\documentclass[prl, aps, 10pt, showpacs, superscriptaddress,
               twocolumn, floatfix]{revtex4-1}
\usepackage{times,graphicx,color,url}
\usepackage[pdftex,bookmarks=false]{hyperref}
\begin{document}
\title{Search for light massive gauge bosons as an explanation of the
  \boldmath{$(g-2)_\mu$} anomaly at MAMI}
\newcommand{\mainz}{\affiliation{Institut f\"{u}r Kernphysik,
    Johannes Gutenberg-Universit\"{a}t Mainz, D-55099 Mainz, Germany}}
\newcommand{\clermont}{\affiliation{ Clermont Universit\'e,
    Universit\'e Blaise Pascal, CNRS/IN2P3, LPC, BP 10448, F-63000
    Clermont-Ferrand, France}}
\newcommand{\zagreb}{\affiliation{Department of Physics, 
    University of Zagreb, HR-10002 Zagreb, Croatia}}
\newcommand{\stefan}{\affiliation{Jo\v{z}ef Stefan
    Institute, SI-1000 Ljubljana, Slovenia}}
\newcommand{\ljubljana}{\affiliation{ Department of Physics,
    University of Ljubljana, SI-1000 Ljubljana, Slovenia}}
\newcommand{\telaviv}{\affiliation{Racah Institute of Physics, Hebrew
    University of Jerusalem, Jerusalem 91904, Israel}}
\author{H.~Merkel}
\thanks{merkel@kph.uni-mainz.de}
\mainz
\homepage{http://www.kph.uni-mainz.de}
\author{P.~Achenbach}
\author{C.~Ayerbe~Gayoso}
\author{T.~Beranek}
\mainz
\author{J.~Beri\v{c}i\v{c}}
\stefan
\author{J.\,C.~Bernauer}
\thanks{Present address: MIT-LNS, Cambridge, MA, USA.}
\author{R.~B\"{o}hm}
\mainz
\author{D.~Bosnar}
\zagreb
\author{L.~Correa}
\mainz
\author{L.~Debenjak}
\stefan
\author{A.~Denig}
\author{M.\,O.~Distler}
\author{A.~Esser}
\mainz
\author{H.~Fonvieille}
\clermont
\author{I.~Fri\v{s}\v{c}i\'{c}}\zagreb
\author{M.~G\'omez~Rodr\'iguez de la Paz}
\author{M.~Hoek}
\author{S.~Kegel}
\author{Y.~Kohl}
\author{D.\,G.~Middleton}
\author{M.~Mihovilovi\v{c}}
\author{U.~M\"{u}ller}
\author{L.~Nungesser}
\author{J.~Pochodzalla}
\author{M.~Rohrbeck}
\mainz
\author{G.~Ron}
\telaviv
\author{S.~S\'{a}nchez Majos}
\author{B.\,S.~Schlimme}
\author{M.~Schoth}
\author{F.~Schulz}
\author{C.~Sfienti}
\mainz
\author{S.~\v{S}irca}
\stefan
\ljubljana
\author{M.~Thiel}
\author{A.~Tyukin}
\author{A.~Weber}
\author{M.~Weinriefer}
\mainz
\collaboration{A1 Collaboration}
\date{April 22, 2014}
\noaffiliation
\begin{abstract}
  A massive, but light abelian $U(1)$ gauge boson is a well motivated
  possible signature of physics beyond the Standard Model of particle
  physics. In this paper, the search for the signal of such a $U(1)$
  gauge boson in electron-positron pair-production at the spectrometer
  setup of the A1 Collaboration at the Mainz Microtron (MAMI) is
  described. Exclusion limits in the mass range of
  $40~\mathrm{MeV}/c^2$ up to $300~\mathrm{MeV}/c^2$ with a
  sensitivity in the mixing parameter of down to $\epsilon^2=8\times
  10^{-7}$ are presented. A large fraction of the parameter space has
  been excluded where the discrepancy of the measured anomalous
  magnetic moment of the muon with theory might be explained by an
  additional $U(1)$ gauge boson.
\end{abstract}
\maketitle
%

\section{Introduction}

The completion of the Standard Model (SM) of particle physics by the
discovery of the Higgs particle at the Large Hadron Collider (LHC) is
undoubtedly a remarkable success after decades of particle physics
experiments~\cite{Chatrchyan201230}. This success, however, also
emphasizes one of the unresolved major questions of today's physics:
the meanwhile well-established existence of dark matter in the
universe is one of the most pressing indications for the need for new
physics beyond the SM. Since there are up to now no experimental hints
from LHC for super-symmetry, which for decades provided the most
promising candidate from particle physics for Dark Matter, the search
for physics beyond the SM has to be extended to more general concepts.

Given the rich structure of the SM it would not be surprising to have
a similar rich structure for a possible \textit{dark sector},
consisting of particles and interactions with only tiny interaction
with SM matter and fields. Most extensions of the SM, like
\textit{e.g.} string theory, provide such a rich structure, which has
to be broken down to the observed SM.

In the last years, a particularly well motivated portal to such a dark
sector, the search for a massive $U(1)$ gauge boson, triggered a vast
amount of theoretical and experimental activities. Such a $U(1)$ gauge
boson, sometimes called ``dark photon'' or $\gamma'$, arises naturally
in several extensions of the SM as lowest rank interaction of this
sector (see \textit{e.g.} ref.~\cite{Langacker:2008yv} for an
overview).

\begin{figure}[b]
\centerline{\includegraphics[width=0.7\columnwidth]{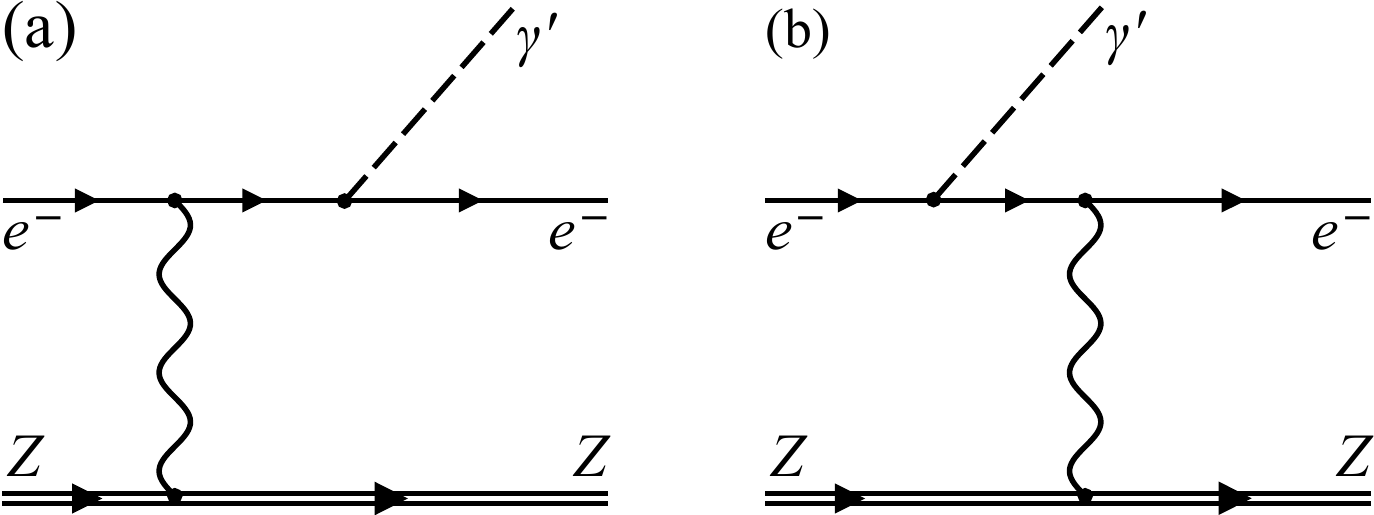}}
\caption{Radiative production of a $\gamma'$ in final (a) and initial
  state (b) on a heavy target nucleus $Z$. The subsequent decay of the
  $\gamma'$ to an electron-positron pair would be the unique signal of
  such a $\gamma'$ with a sharp mass distribution.}
\label{graphs}
\end{figure}

The residual interaction of a dark photon with SM matter is given in a
simplest model by kinematic
mixing~\cite{Holdom:1985ag,Galison:1983pa}, producing an effective
interaction $\epsilon e A_\mu'J^\mu_{\mathrm{EM}}$ of the dark photon
field $A'$ with the electric current $J_{\mathrm{EM}}$. The strength
of this interaction is given by the mixing parameter
$\epsilon^2=\alpha'/\alpha$ equal to the ratio of dark and SM
electromagnetic couplings, and which is not required to be small from
first principles. Assuming that $\epsilon$ vanishes at high energies,
$\epsilon$ can be generated by perturbative corrections including
particles which are charged both under electromagnetic interaction and
the $U(1)$ interaction, leading to a natural scale of $\epsilon \sim
10^{-8} - 10 ^{-2}$. Including non-perturbative models, values of
$\epsilon \sim 10^{-12} - 10 ^{-3}$ have been
discussed~\cite{ArkaniHamed:2008qp,Essig:2009nc}.

Besides the strong motivation from models and theory, several
experimental phenomena could be explained by such a dark photon. A
dark sector with an annihilation channel to dark photons could explain
\textit{e.g.} the positron excess in the universe measured first by
PAMELA~\cite{Adriani:2008zr} and later confirmed by
FERMI~\cite{FermiLAT:2011ab} and AMS-02~\cite{Aguilar:2013qda}. While
other positron sources, for example Quasars, are also discussed in the
literature, the dark photon annihilation process provides a good fit
to the positron spectrum.
 
Of special interest for the parameter range probed in this experiment
is the discrepancy of the measured anomalous magnetic moment $(g-2)$
of the muon~\cite{PhysRevD.73.072003} in comparison with SM
calculations~\cite{Davier:2010nc}. This discrepancy could be explained
by loop contributions of dark photons with a mass range of
$10\,\mathrm{MeV}/c^2-200\,\mathrm{MeV}/c^2$ and a mixing parameter
around $\epsilon^2\approx 10^{-5}$~\cite{Pospelov:2008zw}.

This paper describes the search for a dark photon in the mass region
of $40\,\mathrm{MeV}/c^2-300\,\mathrm{MeV}/c^2$ by a fixed target
electron scattering experiment. A possible dark photon could be
produced radiatively on a heavy target nucleus with high $Z$ (see
Fig.~\ref{graphs}), followed by a subsequent decay into an
electron-positron pair~\cite{Bjorken:2009mm}. Since this decay is
suppressed by the small mixing parameter $\epsilon^2$, the decay width
would be far below the experimental resolution, resulting in a sharp
peak in the invariant mass of the produced lepton pair.

This peak is expected to be on top of a smooth background of standard
radiative electron-positron pair production via a virtual photon. This
background can be calculated in QED and the tools to integrate the
background and a possible signal over the acceptance of the experiment
were developed and discussed in detail in
refs.~\cite{PhysRevD.88.015032,PhysRevD.89.055006}.

\section{Experiment}

The experiment was performed at the spectrometer setup of the A1
Collaboration at the Mainz Microtron (see ref.~\cite{Blomqvist:1998xn}
for a detailed description). The experimental technique was similar to
the technique used in the precursor experiment
\cite{PhysRevLett.106.251802} with a few modifications of the target
and of the vacuum system to further reduce multiple scattering and to
improve the overall mass resolution.

\begin{table*}
  \caption{Kinematical settings. All settings were centered around the
    production of lepton-pairs in beam direction and with maximum energy
    transferred to the pair.}
\label{settings}
\begin{tabular*}{\textwidth}{@{\extracolsep{\fill}}rrrrrrrccccc}
\hline
\hline
Setting 
& Central mass
& \multicolumn{1}{c}{Beam Energy $E_0$} 
& \multicolumn{1}{c}{$\theta_{e+}$} 
& \multicolumn{1}{c}{$p_{e+}$}
& \multicolumn{1}{c}{$\theta_{e-}$} 
& \multicolumn{1}{c}{$p_{e-}$} 
& $e^+$ in Spec.
& $e^-$ in Spec.
& Collimator A 
& Collimator B 
& Target \\
& \multicolumn{1}{c}{$(\mathrm{MeV}/c^2)$}
& \multicolumn{1}{c}{$(\mathrm{MeV})$} & 
& \multicolumn{1}{c}{$(\mathrm{MeV}/c)$} & 
& \multicolumn{1}{c}{$(\mathrm{MeV}/c)$} &&&&&\\
\hline
 1 &  54 & 180 & 20.0$^\circ$ &  74.0 & 15.1$^\circ$ &  97.1 & A & B & 28 msr & 5.6 msr & single foil \\
 2 &  54 & 180 & 15.1$^\circ$ & 100.3 & 20.0$^\circ$ &  74.0 & B & A & 21 msr & 5.6 msr & single foil \\
 3 &  57 & 180 & 20.0$^\circ$ &  78.7 & 15.6$^\circ$ &  98.0 & A & B & 21 msr & 5.6 msr & single foil \\
 4 &  72 & 240 & 20.0$^\circ$ & 103.6 & 15.6$^\circ$ & 132.0 & A & B & 21 msr & 5.6 msr & single foil \\
 5 &  76 & 255 & 20.0$^\circ$ & 105.0 & 15.1$^\circ$ & 137.3 & A & B & 28 msr & 5.6 msr & single foil \\
 6 &  77 & 255 & 20.0$^\circ$ & 110.1 & 15.6$^\circ$ & 140.4 & A & B & 21 msr & 5.6 msr & single foil \\
 7 &  91 & 300 & 20.0$^\circ$ & 129.5 & 15.6$^\circ$ & 164.6 & A & B & 21 msr & 5.6 msr & single foil \\
 8 & 103 & 345 & 20.0$^\circ$ & 142.0 & 15.1$^\circ$ & 186.5 & A & B & 28 msr & 5.6 msr & foil stack \\
 9 & 109 & 360 & 20.0$^\circ$ & 155.4 & 15.6$^\circ$ & 197.6 & A & B & 21 msr & 5.6 msr & single foil \\
10 & 135 & 450 & 20.0$^\circ$ & 185.0 & 15.1$^\circ$ & 243.3 & A & B & 28 msr & 5.6 msr & foil stack \\
11 & 138 & 435 & 15.6$^\circ$ & 244.0 & 20.0$^\circ$ & 190.7 & B & A & 21 msr & 5.6 msr & single foil \\
12 & 138 & 435 & 15.6$^\circ$ & 233.9 & 20.0$^\circ$ & 190.0 & B & A & 21 msr & 5.6 msr & single foil \\
13 & 138 & 435 & 20.0$^\circ$ & 190.0 & 15.6$^\circ$ & 244.5 & A & B & 21 msr & 5.6 msr & single foil \\
14 & 138 & 435 & 20.0$^\circ$ & 190.0 & 15.6$^\circ$ & 234.1 & A & B & 21 msr & 5.6 msr & single foil \\
15 & 150 & 495 & 20.0$^\circ$ & 213.7 & 15.6$^\circ$ & 271.1 & A & B & 21 msr & 5.6 msr & foil stack \\
16 & 170 & 570 & 20.0$^\circ$ & 234.0 & 15.1$^\circ$ & 307.3 & A & B & 28 msr & 5.6 msr & foil stack \\
17 & 177 & 585 & 20.0$^\circ$ & 250.0 & 15.6$^\circ$ & 317.3 & A & B & 21 msr & 5.6 msr & foil stack \\
18 & 202 & 675 & 15.1$^\circ$ & 367.0 & 20.0$^\circ$ & 277.2 & B & A & 21 msr & 5.6 msr & single foil \\
19 & 218 & 720 & 20.0$^\circ$ & 309.2 & 15.6$^\circ$ & 392.7 & A & B & 21 msr & 5.6 msr & foil stack \\
20 & 256 & 855 & 20.0$^\circ$ & 351.0 & 15.1$^\circ$ & 460.3 & A & B & 28 msr & 5.6 msr & foil stack \\
21 & 270 & 855 & 15.2$^\circ$ & 509.4 & 22.8$^\circ$ & 346.3 & B & A & 21 msr & 5.6 msr & single foil \\
22 & 270 & 855 & 15.1$^\circ$ & 511.7 & 20.0$^\circ$ & 346.3 & B & A & 21 msr & 5.6 msr & single foil \\
\hline
\hline
\end{tabular*}
\end{table*}

Table \ref{settings} summarizes the kinematical settings. For all
settings, the incoming electron beam of the accelerator hits a target
consisting of one or several strips of tantalum foils (99.99\%
$^{181}\mathrm{Ta}$) with the thickness of each separate foil between
$1~\mu\mathrm{m}$ and $6~\mu\mathrm{m}$. The target configuration was
optimized separately for each setting for maximum luminosity with
minimized load by radiation background in the focal plane detectors of
the spectrometers.

For the detection of the lepton pair from the decay of a possible dark
photon, the spectrometers A and B of the A1 setup were placed at their
minimal angle (see table \ref{settings}). With these fixed angles, the
settings were adjusted to cover the production of a dark photon in
beam direction and to cover the maximum energy transfer to the dark
photon. The choice of the polarity of the spectrometers was given by
the background conditions. Since the theoretical description of the
background process improved during the analysis, it turns out that
some of these settings were not chosen optimally. The settings of the
pilot experiment~\cite{PhysRevLett.106.251802} were included and
reanalyzed with additional event samples covering the same mass
region.

The vacuum system of the spectrometers was connected with the
scattering chamber to minimize multiple scattering. Both spectrometers
were equipped with four layers of vertical drift chambers for position
resolution, two layers of scintillator detectors for trigger and
timing purposes, and gas \v{C}erenkov detectors for pion-electron
separation and further background reduction.

The beam current of up to $I=80~\mu\mathrm{A}$ was measured with a
fluxgate-magnetometer (F\"orster probe). The angular acceptances of
the spectrometers were defined by heavy metal collimators. For
spectrometer B, a collimator setting of
$40~\mathrm{mrad}~\mathrm{(horizontal)} \times
140~\mathrm{mrad}~\mathrm{(vertical)} = 5.6~\mathrm{msr}$ was used for
all settings, while for spectrometer A two different collimators with
$150~\mathrm{mrad} \times 140~\mathrm{mrad} = 21~\mathrm{msr}$ and
$200~\mathrm{mrad} \times 140~\mathrm{mrad} = 28~\mathrm{msr}$ were
used. The momentum acceptance of the spectrometers was $20\%$ for
spectrometer A and $15\%$ for spectrometer B.

\begin{figure}
\includegraphics[width=\columnwidth]{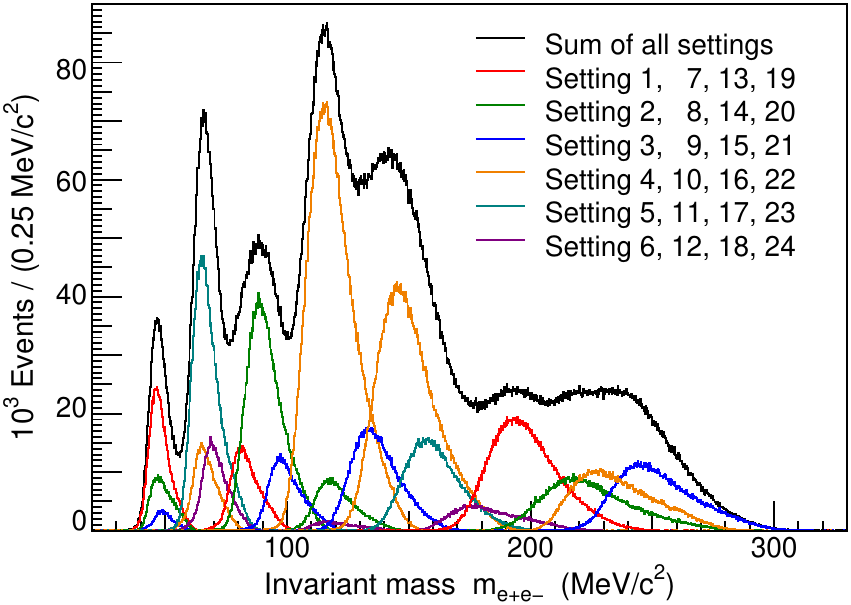}
\caption{(color online). ~~Mass distribution of the individual
  settings (color/shaded) and of the sum (black). The experiment
  probes the invariant mass region between 40 and
  $300~\mathrm{MeV}/c^2$.}
\label{allsettings}
\end{figure}

\section{Data Analysis}

The lepton pair was detected in coincidence between the two
spectrometers. For reaction identification, a cut was applied first on
a signal in the \v{C}erenkov detectors of both spectrometers with an
efficiency of $\approx 98\%$. The coincidence time between
spectrometer A and B was corrected for the path length in spectrometer
A of $\approx 10~\mathrm{m}$ and spectrometer B of $\approx
12~\mathrm{m}$. After this correction a clear coincidence peak with a
width of less than $1~\mathrm{ns}$ (FWHM) was seen. The range of
$|\Delta t_{A\and B}|<1~\mathrm{ns}$ was used to identify lepton
pairs. The background contribution from random coincidences was
estimated by a cut on the sideband with $5~\mathrm{ns}<|\Delta
t_{A\and B}|<15~\mathrm{ns}$.

Additional cuts were applied for the acceptance of the spectrometers
to further reduce the contribution of backscattered particles from the
entrance flange of spectrometer B. Finally, cuts on the validity of
the overall kinematics was applied to remove \textit{e.g.} accidental
coincidences where the total energy of the pair exceeds the beam
energy.

In total the background contribution ranges from $4\%$ up to $11\%$
after all cuts. This background contribution is not subtracted for the
peak search but has to be taken into account later in the calculation
of the exclusion limit.

For the identified lepton pairs, the invariant pair mass was
determined by the four-momenta of the leptons via $m_{\mathrm{e+e-}}^2
= (p_{e^+} + p_{e^-})^2$.  Fig. \ref{allsettings} shows the mass
distribution of all settings.

To add up the pair mass distribution of all settings, the absolute
mass calibration of each setting has to be better than the expected
peak width. The magnetic field of the spectrometers was monitored with
NMR-probes to $\delta B/B = 10^{-4}$ and simultaneously with
Hall-probes on the $\delta B/B = 5\times 10^{-4}$ level. This
translates in total to a mass calibration of better than
$100~\mathrm{keV}/c^2$. The calibration was verified at several points
by additional measurements of elastic scattering on tantalum. The
position and width of the $^{181}\mathrm{Ta}$ ground state was used to
confirm the total calibration and to extract the momentum and angular
resolution of the total setup \textit{in situ}. The experimental
resolutions were used to tune the detailed simulation of the elastic
scattering process to reproduce the elastic peak shape. Finally the
simulation was used to determine the mass resolution and expected dark
photon peak shape in dependence of the mass including radiative
corrections. The resulting resolution varies between
$210~\mathrm{keV}/c^2$ FWHM in the lowest mass range up to
$920~\mathrm{keV}/c^2$ FWHM for the settings of the last experiment.

The estimated peak shape was used to perform a search for a peak in
the total mass distribution. For this the background for each bin was
estimated by a local fit of the neighboring bins with a cubic
polynomial. The confidence interval was determined using the
Feldman-Cousins algorithm~\cite{Feldman:1997qc} (please note that in
the literature several different approaches were used by different
experiments to determine limits for dark-photon searches, they differ
however only by a few percent). The results were corrected for the
leakage of the peak outside the bin. The complete procedure was
repeated with shifted binning limits in eight steps.

No significant signal for a dark photon was detected.

\section{Results and Interpretation}

Due to the use of thin tantalum foil stacks as targets the
normalization of the cross section contains large
uncertainties. However the identification of the QED background
process is very clean and can be used as normalization. Therefore, to
translate the exclusion limit in terms of events to an exclusion limit
in terms of the mixing parameter $\epsilon$ we used the ratio of dark
photon production with mixing parameter $\epsilon$ divided by the QED
background process~\cite{Bjorken:2009mm}
\[
R= \frac{d\sigma({X\rightarrow \gamma'\,Y\rightarrow e^+e^- Y})}{
  d\sigma({X\rightarrow \gamma^* Y\rightarrow e^+e^- Y})} =
\frac{3\pi}{2N_{\mathrm{f}}} \frac{\epsilon^2}{\alpha}
\frac{m_{\gamma'}}{\delta_m}.
\]
Here $N_{\mathrm{f}}$ is the ratio of the phase space of the decay
into an $e^+e^-$ pair to the phase space of the total decay (equal to
1 below $2m_\mu$) and $\delta_m$ is the bin width in mass. For the
virtual photon channel we used the background-subtracted mass
distribution. To determine the ratio $R$ both cross sections as
calculated in ref.~\cite{PhysRevD.88.015032} were integrated over the
acceptance of the experiment by standard Monte-Carlo methods. Here,
the normalization was chosen to reproduce the measured mass
distribution.

Please note that in the interpretation of the data in
ref.~\cite{PhysRevLett.106.251802} the cross sections were calculated
not including the full anti-symmetrization as discussed in
ref.~\cite{PhysRevD.88.015032}, leading to an overestimation of the
sensitivity by a factor 2--3. Therefore these data were included in
this analysis and reanalyzed. Since additional data were taken in the
same mass range, roughly the same sensitivity was achieved.

\begin{figure}
\includegraphics[width=\columnwidth]{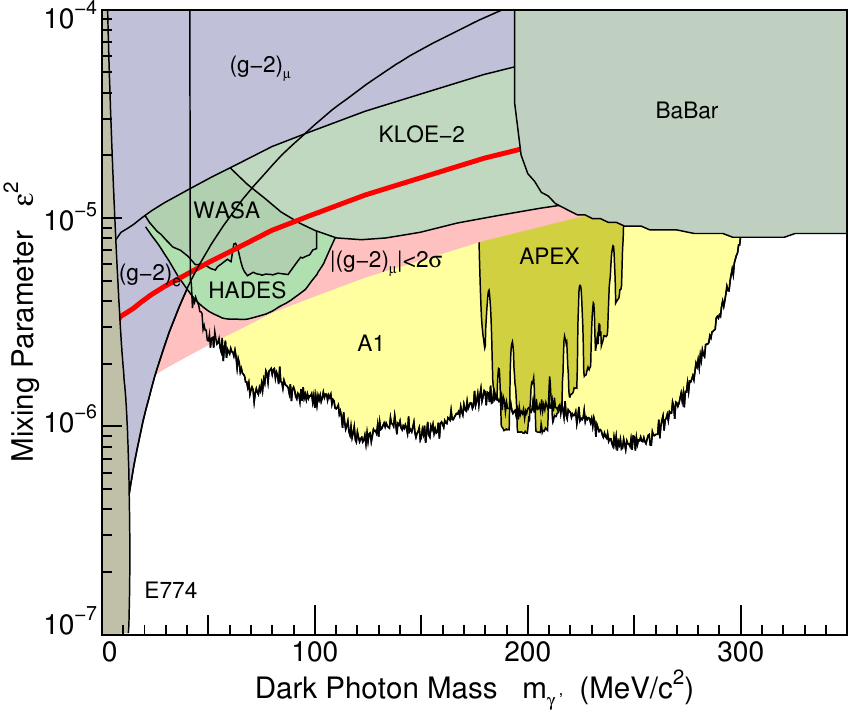}
\caption{(color online). ~~Exclusion limits in terms of mixing
  parameter $\epsilon$. The yellow (light shaded) area marked with A1
  is excluded by this experiment.}
\label{exclusionlimits}
\end{figure}

Fig.~\ref{exclusionlimits} shows the resulting $2\sigma$ exclusion
limits. Also included in the figure are the limits by the APEX
\cite{PhysRevLett.107.191804}, WASA-at-COSY~\cite{Adlarson}, KLOE-2
\cite{Babusci:2012cr}, HADES~\cite{Agakishiev2014265}, and BaBar
\cite{Aubert:2009cp,Echenard} collaborations. The red line shows the
interpretation of the $(g-2)_\mu$ discrepancy as a dark photon with a
$2\sigma$ error band (red shaded region) and as exclusion limit (blue
shaded region). Also included is the reanalysis of
ref.~\cite{PhysRevD.86.095019} of the beam dump experiment
E774~\cite{PhysRevLett.67.2942} to extract exclusion limits for dark
photons.

With the new measurement presented here, the exclusion limit in the
region of the $(g-2)$ anomaly of the muon was improved
considerably. While the results of the meson decays by KLOE-2,
WASA-at-COSY, and HADES were not able to completely rule out the dark
photon as the origin of the anomaly, the new data set clearly covers
the possible signal of the anomaly by several sigmas over a large mass
range. The remaining undecided mass range of $25\,\mathrm{MeV}/c^2
\lesssim m_{\gamma'} \lesssim 50\,\mathrm{MeV}/c^2$ cannot be covered
by the spectrometers of the A1 collaboration without
modifications. However, several experiments by different
collaborations are already planned to access the low mass region in
the near future (see ref.~\cite{PhysRevD.89.055006} for a summary).

\paragraph{Acknowledgments} 

The authors like to thank the MAMI accelerator group for the excellent
beam quality which made this experiment possible. This work was
supported by the Collaborative Research Center 1044 and the State of
Rhineland-Palatinate.
\bibliographystyle{apsrev4-1} 
\bibliography{DarkPhoton2013}
\end{document}